\documentstyle[11pt,epsfig]{article}
\newcommand{\be}{\begin{equation}}
\newcommand{\ee}{\end{equation}}
\newcommand{\ba}{\begin{eqnarray}}
\newcommand{\ea}{\end{eqnarray}}
\newcommand{\no}{\nonumber \\}
\newcommand{\ep}{\epsilon}
\newcommand{\vk}{{\bf k}}
\def\la{\langle}\def\ra{\rangle}
\def\Tr{\rm Tr}
\begin{document}
\begin{titlepage}
\begin{flushright}
KIAS-P99051
\end{flushright}
\begin{center}
{\Large\bf Competition Between Induced Symmetry Breaking, Cooper
Pairing and Chiral Condensate at Finite Density}

\vskip 0.3in {Youngman Kim$^{a,c,d}$ and Mannque Rho$^{b,c,d}$ }\\
\vskip 0.3in {\it a)  Department of Physics, Hanyang University,} \\
{\it Seoul 133-791, Korea}\\
\vskip 0.1cm
{\it  b)  Service de Physique Th\'{e}orique, CEA  Saclay}\\
{\it 91191 Gif-sur-Yvette Cedex, France}\\
\vskip 0.1cm
{\it  c)  School of Physics, Korea Institute for Advanced Study}\\
{\it Seoul 133-791, Korea }\\
{\it  d)}  {\it Department of Physics, Seoul National University,} \\
{\it Seoul 151-742, Korea}
\vskip 0.3in

{\bf ABSTRACT}\\ \vskip 0.3cm
\end{center}

We study the competition between induced symmetry breaking (ISB),
 Cooper pairing (superconductivity) and chiral
condensation at finite
non-asymptotic density. Using a quantum mechanical
model studied recently by Ilieva and Thirring, we analyze the
expectation value of the fermion number density $\la a^*a\ra$ in
the presence of chemical potential and discuss relevance of
the ISB phenomenon in the model. By Bogoliubov transformation, we obtain the
ground state energy and find that a phase with both
superconducting and mean field gap is a true ground state. We also
study the effective potentials of a two-dimensional field theory
model for competitions between $\la\bar\psi\psi\ra$,
$\la\psi\psi\ra$ and $\la\psi^\dagger\psi\ra$. We find that in the
regime $\mu <\mu_c$, where $\mu_c$ is found to be around $0.4 m_F$
with $m_F$ being the renormalizaton point of chiral condensate, we
have a chiral symmetry-breaking phase with almost zero fermion
number density and in the case of $\mu >\mu_c$ the system sits in
the mixed phase characterized by
 $\la \psi^\dagger\psi\ra\neq 0$ and  $\la \psi\psi\ra\neq 0$
as long as the renormalization point, $M_0$, corresponding to a
superconducting gap is not zero, $M_0\neq 0$.

\end{titlepage}

\section{Introduction}
\indent\indent What happens to QCD at large density is a
fascinating topic which is generating an intense activity in nuclear
and particle communities. It seems that at an asymptotic density,
the likely scenario is that diquarks condense into the matter as a
consequence of which color superconductivity (CSC) may take place.
For a recent review with references, see \cite{CSC}. But it is not
at all clear whether the density involved for CSC \cite{rs} is
relevant to the dense matter that can be accessed by experiments
and that we are interested in, namely neutron stars and heavy ion
collisions that are currently studied or planned for the future.
The question we would like to address here is: What happens to
nuclear matter as density increases beyond the normal matter
density toward the critical density for chiral phase transition?
This question is not easy to answer since the coupling involved is not
weak enough as what might happen at asymptotic density at which
QCD weak-coupling property can be exploited but it is definitely
more relevant to the physics we want to study in conjunction with
experiments. We expect a priori that a variety of different
phenomena will intervene in the regime of density that we are
interested in and these may not render themselves to simple
analyses. Some of them were discussed from a weak-coupling QCD
point of view in \cite{PRWZ}. Here we address them from a
different perspective.

The issue in question was studied by Langfeld and Rho~\cite{lr} using a
semi-realistic field theory model of the NJL type. Some of the
results found there were novel and suggestive but certain
restrictive features of the NJL-type models (i.e., no confinement)
rendered them problematic. See ~\cite{birse} for discussions on
related matters. Here we would like to analyze the problem using a
simplified solvable quantum mechanical model and a two-dimensional
field theory model. Our study here is exploratory and we cannot
make quantitative statements (for instance, our analysis is
``blind" to color and flavor contents) as to what really must
happen but our hope is to gain some qualitative understanding of
what might be taking place generically at a density greater than
that of normal nuclear matter but much less than the asymptotic
density relevant to the color superconductivity.

In \cite{lr}, Langfeld and Rho suggested that when the isoscalar
vector channel (called the $\omega$ channel) becomes sufficiently
attractive as expected in the presence of collective phonon modes
of the pertinent quantum numbers as discussed in ~\cite{KRBR},
then the baryon density can have a significant jump at some
critical chemical potential in conjunction with generating
low-mass excitations with the quantum numbers of the $\omega$
meson. Since this can be expressed as a rapid increase -- but
 not a discontinuous jump -- in the ground state
expectation value of the time component of the $\omega$ field, the
process was called ``induced symmetry breaking" (ISB) associated
with the broken isoscalar vector symmetry characterized by the
``vaccum" expectation value of the time component of the vector
current. We should stress that in medium, $\la\omega_0\ra$ is of
course never zero, being proportional to density. Therefore
$\la\omega_0\ra$ cannot, strictly speaking, be used as a signal
for a phase change. However it can have an anomalous increase at
certain chemical potential reminiscent of an order parameter that
can be associated with a phase transition. A simple condensed
matter model that illustrates this phenomenon is discussed by
Langfeld~\cite{langfeld}. We shall use somewhat abusively
 this notion in this paper with the caveat in mind.

A similar behavior of the baryon density vs. the chemical potential seen
in the ISB has been also observed by Berges et. al~\cite{berg} in
a renormalization-group flow analysis of a linear sigma model in
which it was found that quark number density shows non-analytic
behavior at $\mu  =1.025M_q$, with $M_q\sim 316.2 MeV$ an effective
constituent quark mass,
where the vacuum expectation value
of the $\sigma$-field, chiral condensate, vanishes.
As discussed in ref.\cite{lr}, this is
easy to understand in terms of change in the number distribution
at that $\mu$. The change in the number distribution
can give a jump in the number density through
$\rho\sim\int dk [\exp \beta (\epsilon(k)-\mu)+1]^{-1}$.
That there is a rapid increase in density at the
transition point is intrinsic in the chiral phase transition. The
point of ref.\cite{lr}, however, was that the ISB is concurrent
with chiral symmetry restoration and that the presence of the ISB
could postpone color superconductivity until the $\omega$-channel
becomes ineffective at some high density due to
weakened QCD coupling. At what density this can
happen, we cannot say but the density regime involved could be
relevant to the interior of compact stars.

In this paper, we wish to address the problem in two aspects.
First we take a quantum mechanical model studied by Ilieva and
Thirring (IT)~\cite{it1,it2} and study, exploiting its
solvability, whether and how the ISB manisfests itself in the
model. As stressed by IT, this analysis does not depend upon the
dimension of the space considered. We shall discuss in section 2
 the ISB phenomenon in the IT model
and find that the model can have a phase in
which  both the ISB and fermion pair (Cooper) condensate co-exist
and find that the ground state energy of the mixed phase is indeed
lower than the normal state without such condensates. This model
does not however render itself to a discussion of what happens with chiral
(fermion-antifermion) condensates.  To investigate whether this
kind of mixed phase is possible and also how the chiral condensate
figures in that phase, we resort to a soluble field theory model
in two space-time dimensions studied by Chodos et al~\cite{chodos}
and examine the effective potentials to see whether both
condensates can coexist in the {\it global minimum}. The
two-dimensional model is subject to the
Mermin-Wagner-Coleman~\cite{mwc} no-go theorem for spontaneous
symmetry breaking
but we follow ref.\cite{chodos} in considering
this model in the large $N$ sense a la Witten~\cite{witten}. It
would be possible to formulate the problem in effective two
dimensions dimensionally reduced from four dimensions, thereby
avoiding the Mermin-Wagner-Coleman theorem. We find in section 3
that although gap equations exist in which both condensates are
non-zero, the global minimum of the effective potential always
occurs for the case when one or the other condensate vanishes as
long as $\la\bar\psi\psi\ra$ and $\la\psi^\dagger\psi\ra$ are
concerned. On the other hand, we do find a stable mixed phase
consisting of $\la\psi\psi\ra$ and $\la\psi^\dagger\psi\ra$ at
some high density. Section 4 contains concluding remarks. In
Appendix we clarify the notion of grand canonical ensemble used in
this paper and in \cite{it1}.

\section{Ilieva-Thirring Model and the ISB}
\indent\indent Ilieva and Thirring~\cite{it1} studied the
competition between $\la\psi^*\psi\ra$ and $\la\psi\psi\ra$
in a solvable quantum mechanical model (that we shall
refer to as IT). Here we revisit their discussion in light of our
objective as defined in the previous section. Since we shall use
the notation and the results of IT, we first summarize them and
then extract the information relevant to us.

 It was argued in
\cite{it1}
 that two Hamiltonians are  equivalent if they
lead to the same time evolution of the local observables. This
means that the effective Hamiltonian $H_B$ \ba H_B=\int
dp\{a^*(p)a(p)[\omega (p) +\Delta_M (p)]+\frac{1}{2}\Delta_B (p)
[a^*(p)a^*(-p)+a(-p)a(p)]\}\label{effh1} \ea is equivalent to
$H=H_{kin} +V_B +V_M$, \ba H&=&H_{kin} +V_B +V_M\no
V_B&=&\kappa^{3/2}\int dp dp^\prime dq dq^\prime
a^*(q)a^*(q^\prime) a(p)a(p^\prime)  v_B(q, q^\prime,
p,p^\prime)e^{-\kappa(p+p^\prime)^2-\kappa (q+q^\prime)^2}\no
V_M&=&\kappa^{3/2}\int dp dp^\prime dq dq^\prime
a^*(q)a^*(q^\prime) a(p)a(p^\prime)  v_M(q, q^\prime,
p,p^\prime)e^{-\kappa(q-p)^2-\kappa
(q^\prime-p^\prime)^2}\nonumber \ea where $V_B$ and $V_M$
support respectively the pairing gap $\Delta_B$ and the
``mean-field gap" $\Delta_M$ for $\kappa\rightarrow \infty$,
provided the gap equations
\ba
[a(k) , V_B+V_M ] =[a(k),
H_B-H_{kin}]\label{gap}
\ea
are satisfied. To solve the gap
equations, IT choose the following potential describing an
interaction concentrated about the Fermi surface
\ba &&v_{B,
M}(\vec k,\vec p)=\lambda_{B,M} S(\vec k)S(\vec p)\no {\rm with}~
S(\vec k)&=&\frac{1}{2\epsilon}[\Theta (|\vec
k|-\sqrt{\mu}+\epsilon) -\Theta (|\vec k|-\sqrt{\mu} -\epsilon)]
\ea where $\Theta (x)$ is the Heaviside function. With this
potential and the additional assumption $\omega (p)=p^2=\mu $ (with
$2m=1$), the gap equations read
 \ba
\frac{1}{2}\Delta_M(\mu)&=&\lambda_M \{\frac{c^2(\mu)}{1+e^{\beta
(\bar W-\mu)}} + \frac{s^2(\mu)}{1+e^{-\beta (\bar W-\mu)} } \}
\label{ge1}\\ \bar W&=& \lambda_B\tanh\frac{\beta (\bar W-\mu)}{2}
~~~{\rm or}~~~\Delta_B=0\label{gap2}\\ && \bar
W=\sqrt{[\mu+\Delta_M]^2 +\Delta_B^2}\label{gap1}
 \ea
(where the pairing gap equation can have two solutions
(\ref{gap2}), the former non-trivial and the latter
trivial)
 with the subsidiary conditions
  \ba
c^2(\mu)+s^2 (\mu) &=&1,
\no c^2(\mu)-s^2 (\mu) &=&(\omega (\mu ) +\Delta_M) /
\bar W,\no
 2c(\mu)s (\mu) &=&\Delta_B(\mu )/\bar W
\ea
 where $c$ and $s$
are real coefficients. It follows from (\ref{ge1}) that ${\rm
sgn}\Delta_M={\rm sgn}\lambda_M$. It has been assumed that $\bar W
>0$. From (\ref{effh1}), we have the following thermal expectation
value of the number density operator $a^*a$,
 \ba \la
a^*(p)a(p^\prime)\ra=\delta(p-p^\prime)
\{\frac{c^2(p)}{1+e^{\beta (\bar W-\mu)}} +
\frac{s^2(p)}{1+e^{-\beta (\bar W-\mu)} } \}\label{mf1}. \ea

\subsection{Ground-State Energy }
\indent\indent
 Let us now reformulate the IT
model by performing Bogoliubov transformation explicitly and
evaluate the ground-state energy which is related to the
``effective energy" from which we can derive the same quantities.
We write the reduced Hamitonian in the form
 \ba H=\sum_{\bf
k\sigma }\ep_\vk a_{\vk\sigma}^* a_{\vk\sigma} + \sum_{\bf k\bf l
} V_{\vk\bf l}^B a_\vk^* a_{-\vk}^* a_{-\bf l} a_{\bf l} +
\sum_{\bf k\bf l } V_{\vk\bf l}^M  a_\vk^* a_{ \bf l}^* a_{\vk}
a_{\bf l}.\label{rh1}
 \ea
  Introduce ``order parameters" $d_\vk$ and
$m_\vk$ by writing
 \ba a_{-\vk}a_\vk &=&d_\vk + (a_{-\vk}a_\vk -
d_\vk )\no a_{\vk}^*a_\vk &=&m_\vk +(a_{\vk}^*a_\vk -m_\vk)
 \ea
and define
 \ba \Delta_\vk &\equiv& -\sum_{\bf l} V_{\vk \bf l}^B
d_{\bf l} = -\sum_{\bf l} V_{\vk \bf l}^B \la a_{-\bf l} a_{\bf
l}\ra \no \bar \Delta_\vk &\equiv& -\sum_{\bf l} V_{\vk \bf l}^M
m_{\bf l} = -\sum_{\bf l} V_{\vk \bf l}^M  \la a_{ \bf l}^* a_{\bf
l}\ra. \ea
 In terms of $\Delta_\vk$ and $\bar \Delta_\vk$, the reduced
Hamiltonian (\ref{rh1}) becomes \ba H &=& \sum_{\bf k\sigma
}\ep_\vk a_{\vk\sigma}^* a_{\vk\sigma} -\sum_{\bf k }(\Delta_\vk
a_\vk^* a_{-\vk}^* + \Delta_\vk^* a_{-\vk} a_{\vk}-\Delta_\vk
d_\vk^*  )\no &&+\sum_{\bf k} (2\bar\Delta_\vk  a_\vk^* a_{\vk}
-\bar\Delta_\vk m_\vk) .\label{mh1} \ea
 To diagonalize the
Hamiltonian, we make use of the Bogoliubov-Valatin canonical
transformation~\cite{bogo} with real coefficients
 \ba a_\vk =c_\vk
b_\vk +s_\vk b_{-\vk}^*\no a_{-\vk} =c_\vk b_{-\vk} - s_\vk
b_{\vk}^*\label{bv}. \ea
 Substituting these new operators into
(\ref{mh1}), we obtain \ba H&=&\sum_{\bf k }\ep_\vk [(c_\vk^2-
s_\vk^2)(b_\vk^* b_\vk +b_{-\vk}^* b_{-\vk}) +2c_\vk s_\vk b_\vk^*
b_{-\vk}^*+ 2c_\vk s_\vk b_{-\vk} b_{\vk} +2s_\vk^2]\no
&&+\sum_{\bf k }[c_\vk s_\vk (\Delta_\vk +\Delta_\vk^*
)(b_\vk^*b_\vk +b_{-\vk}^* b_{-\vk} -1) -(\Delta_\vk
c_\vk^2-\Delta_\vk^* s_\vk^2)b_\vk^* b_{-\vk}^*\no &&+(\Delta_\vk
s_\vk^2-\Delta_\vk^* c_\vk^2)b_{-\vk} b_{\vk} +\Delta_\vk d_\vk^*]
 +\sum_{\bf k }[2\bar\Delta_\vk (c_\vk^2 b_\vk^* b_\vk
+c_\vk s_\vk b_\vk^* b_{-\vk}^*\no
&&+c_\vk s_\vk b_{-\vk} b_{\vk} +s^2 b_{-\vk} b_{-\vk}^*) -\bar\Delta_\vk m_\vk].\label{hf}
\ea
 The diagonalization is effected by demanding that the coefficients of
 $b_\vk^* b_{-\vk}^*$ and $b_{-\vk} b_{\vk}$ vanish
\ba 2c_\vk s_\vk\ep_\vk +\Delta_\vk^*s_\vk^2 -\Delta_\vk c_\vk^2 +
2\bar\Delta_\vk c_\vk s_\vk =0. \ea Multiplying
$\Delta_\vk^*/c_\vk^2$ and defining
$x=\frac{s_\vk}{c_\vk}\Delta_\vk^*$, we get two solutions
 \ba
x=-(\ep_\vk +\bar\Delta_\vk)\pm\sqrt{(\ep_\vk +\bar\Delta_\vk)^2+
\Delta_\vk^2}. \ea
 We take ({\bf +}) sign here to get a stable
minimum energy solution. Then assuming $\Delta_\vk$ to be a real
quantity, we have two equations for $c_\vk$ and $s_\vk$:
 \ba
\frac{s_\vk}{c_\vk} &=&\frac{E_\vk-(\ep_\vk
+\bar\Delta_\vk)}{\Delta_\vk}\no c_\vk^2 + s_\vk^2&=&1 \ea where
$E_\vk\equiv \sqrt{(\ep_\vk +\bar\Delta_\vk)^2+\Delta_\vk^2}$.
Solving the equations, we have \ba c_\vk^2
=\frac{1}{2}(1+\frac{\ep_\vk +\bar\Delta_\vk}{E_\vk} )\no
 s_\vk^2=\frac{1}{2}(1-\frac{\ep_\vk +\bar\Delta_\vk}{E_\vk} ).\label{cs1}
\ea
 To obtain the (coupled) gap equations for $\Delta_\vk$ and
$\bar\Delta_\vk$, we rewrite $\Delta_\vk$ and $\bar\Delta_\vk$ in
terms of the new operators defined in (\ref{bv}), i.e.,
 \ba
\Delta_\vk &=& - \sum_{\bf l} V_{\bf k l}^B c_{\bf l} s_{\bf l}
\la 1 -b_{\bf l}^* b_{\bf l}
 - b_{-\bf l}^* b_{-\bf l}\ra\no
&=& -\sum_{\bf l} V_{\bf kl}^B\frac{\Delta_{\bf l}}{2E_{\bf l}}
\tanh \frac{\beta(E_{\bf l}-\mu)}{2}\no \bar\Delta_\vk &=& -
\sum_{\bf l} V_{\bf kl}^M \la s_{\bf l}^2 +c_{\bf l}^2 b_{\bf l}^*
b_{\bf l} - s_{\bf l}^2 b_{\bf l}^* b_{\bf l}\ra \no
 &=& - \sum_{\bf l} V_{\bf kl}^M
[\frac{c_{\bf l}^2}{e^{\beta(E_{\bf l}-\mu)} +1} + \frac{s_{\bf
l}^2}{e^{-\beta(E_{\bf l}-\mu)} +1}] \label{gape1} \ea
 where we have used $\la b_{\pm \bf l}^* b_{\pm \bf l}\ra=1/(e^{\beta(E_{\bf
l}-\mu)} +1)$. To compare (\ref{gape1}) with the gap equations of
IT model, we take $V_{\bf kl}^B=-2\lambda_B \delta_{|{\bf
k}|,\sqrt{\mu}}\delta_{|{\bf l}|,\sqrt{\mu}}$,
$V^M=-2\lambda_M\delta_{|{\bf k}|,\sqrt{\mu}}\delta_{|{\bf
l}|,\sqrt{\mu}}$ with $\delta_{a,b}$ being Kronecker delta,
$\Delta_\vk =\Delta_B(\mu )$ and $\bar\Delta_\vk =\Delta_M({\mu}
)$ and get \ba E (\mu) &=& \lambda_B \tanh \frac{\beta(E
(\mu)-\mu)}{2} ~or~\Delta_B(\mu )=0\no \Delta_M (\mu) &=&
 2\lambda_M [\frac{c^2(\mu)}{e^{\beta(E(\mu)-\mu)} +1} + \frac{s^2(\mu)}{e^{-\beta(E(\mu)-\mu)} +1}].
\ea
 We can easily read off the ground state energy $U$ from
(\ref{hf}), \ba U&=&\sum_{\bf k }[2\ep_\vk s_\vk^2 -2\Delta_\vk
c_\vk s_\vk +\Delta_\vk d_\vk^*- \bar\Delta_\vk m_\vk
+2\bar\Delta_\vk s_\vk^2]\no &=& \mu - E (\mu)   +\Delta_M (\mu)
+\frac{\Delta_B^2(\mu)}{2\lambda_B} -
\frac{\Delta_M^2(\mu)}{2\lambda_M} \ea where the summation has
been performed with $\delta_{|{\bf k}|,\sqrt{\mu}}$ in the spirit
of the IT model\footnote{In the usual BCS theory, the summation is
replaced by the integration, $\sum_{ k }\approx \int_0^{\hbar
\omega_c}$ with $\hbar \omega_c$, a typical phonon energy.
But since we are considering the very special case of an
interaction concentrated about the Fermi surface, we impose the
Kronecker delta in the summation.}. We are now in a position to
analyze various cases of interest.

\subsection{ $\Delta_B=0$}
\indent\indent
 This case corresponds to the simplest solution of
 gap equations for {\it all} values of $\lambda_M,
\lambda_B,\mu$.
 \vskip 0.5cm

 $\odot$ Case $I$-i : $\mu+\Delta_M
>0$\\ In this case,
 \ba E-\mu=\Delta_M& =
&\frac{2\lambda_M}{1+e^{\beta\Delta_M}}\no \Delta_M &\rightarrow &
\{ \begin{array}{cc} \            2\lambda_M          &   {\rm
for}~ \lambda_M <0       \\
             0               &   {\rm for} ~ \lambda_M >0
\end{array}  ,
\ea as ${\beta \rightarrow \infty}$. Since the phase with
$\Delta_M = 2\lambda_M $ has a non-zero positive ground state
energy $U= -2\lambda_M $ for $ \lambda_M <0$, it does not interest
us. We will not consider it anymore. Then, (\ref{mf1}) becomes
(after integrating over $p^\prime$)
 \ba
\la a^*(p)a(p)\ra &=&\frac{1}{1+e^{\beta (\ep -\mu)}} ~~ {\rm for} ~
\Delta_M =0 \ea where $\ep=p^2$. Note that this phase is nothing
but the normal state (
 $ \Delta_B=0~{\rm and}~ \Delta_M =0$).

\vskip 0.5cm
 $\odot$ Case $I$-ii : $\mu+\Delta_M <0$\\
To satisfy
the condition  $\mu+\Delta_M <0$, $\Delta_M$ should be negative
and therefore $\lambda_M <0$.
 Noting that  $(\ep_\vk +{\bar\Delta}_\vk)/E_\vk =-1$ when we take
$\Delta_\vk (\Delta_B)=0$ and $\ep_\vk =\mu$ in (\ref{cs1}),
 we have $c^2(\mu)=0$ and $s^2(\mu)=1$.
 Therefore we have
 \ba E+\mu&=&-\Delta_M \no
\Delta_M&=&\frac{2\lambda_M}{1+e^{\beta(\Delta_M+2\mu)}}\no
&\rightarrow & \{ \begin{array}{cc} \            2\lambda_M & {\rm
for}~ \Delta_M+2\mu <0       \\
             0               &   {\rm for} ~ \Delta_M+2\mu >0
\end{array}
\ea
as ${\beta \rightarrow \infty}$.
Noting that the second solution $\Delta_M =0$ cannot satisfy the condition
 $\mu+\Delta_M <0$, we have at a given energy $\ep (p)=p^2$
\ba
\la a^*(p)a(p)\ra &=& \frac{1}{1+e^{\beta(\mu-|\ep +\Delta_M |
) } } ~,~ \ep +\Delta_M<0 ,\no
or &=&  \frac{1}{1+e^{\beta(\ep +\Delta_M -\mu
) } } ~,~ \ep +\Delta_M>0\label{sibs}
\ea
 for $\Delta_M =  2\lambda_M$.
 The corresponding ground state energy  $U=2(\mu+\lambda_M)$
is  negative as long as $\mu <
|\lambda_M |$ and therefore this phase is energetically favorable
compared with the normal state of matter.
Since the $\Delta_M$ in (\ref{sibs}) corresponds to $\la V_0\ra$ in eq.(2)
of ref.\cite{lr},
one may be tempted to conclude that
it is a signal of an ISB. But this can not be a candidate for an ISB
because $\Delta_M$ in the IT model cannot be associated with
spontaneous breaking of a symmetry and furthermore $\Delta_M$ is a constant
independent of the chemical potential.

\subsection{ $\Delta_B\neq 0$, $\lambda_B <0 $}
 \indent\indent
In this case, we cannot expect the Cooper pairs to condensate if
 $\lambda_M =0$ as one can see clearly from the gap equation (\ref{gap2}).
If we take $\lambda_M =0$ and therefore
$\Delta_M =0$, the gap equation becomes \ba \sqrt{\mu^2
+\Delta_B^2} = \lambda_B ~{\rm for}~ \Delta_B\neq 0 ~{\rm and}~
\beta\rightarrow\infty \ea which is contradictory with the
conditions we are starting with.

It follows that the parameters under consideration must respect
the following conditions~\cite{it1}:
\ba
 \lambda_B < 0,~ E <\mu,~\Delta_M <0,~\lambda_M<0.
\ea
Since we are interested in the physics of cold dense matter, we
solve the gap equation taking ${\beta \rightarrow \infty}$ and
get~\cite{it1,it2}
\ba
E&=&-\lambda_B,~\Delta_M=\lambda_M\frac{\lambda_B-\mu}{\lambda_B+\lambda_M}\no
\Delta_B&=&\pm\frac{\lambda_B}{\lambda_B+\lambda_M}\sqrt{(\lambda_B-\mu)
(\lambda_B+2\lambda_M+\mu)}\ .\label{sols}
\ea
Note that we have
solutions only when $\mu>E=-\lambda_B$.
To investigate the
competition between the mixed phase given by
(\ref{sols}) and the $\Delta_M\neq 0$ phase in
$I$-ii, we take the following values of coupling constants
with arbitrary dimension
\ba
S_1&:& \lambda_B =-5 ,~\lambda_M =-5\no S_2 &:&
\lambda_B =-5,~\lambda_M =-10\no
S_3 &:& \lambda_B =-5 ,~\lambda_M
=-2\label{cc1}
\ea
and find the following critical chemical
potential $\mu_c$
\ba S_1 &:& \mu_c=5\no S_2 &:&
\mu_c=7.6 \no
S_3 &:& \mu_c^1=2,~~
\mu_c^2=5.
\ea
 In the case of $S_1$ and $S_2$, our
system sits in the phase with $\Delta_B=0$ and $\Delta_M\neq 0$,
$I$-ii, in the regime $\mu <\mu_c$ and rolls down to $\Delta_B
\neq 0$ and $\Delta_M\neq 0$ above $\mu_c$. The case $S_3$ is
quite puzzling in the sense that in the region  $\mu <\mu_c^1$ our
system sits in the phase with $\Delta_B=0$ and $\Delta_M\neq 0$,
for $\mu_c^1<\mu <\mu_c^2$, we have the normal phase ($\Delta_B=0$
and $\Delta_M= 0$) and in the region  $\mu >\mu_c^2$, our system
sits in the $\Delta_B\neq 0$ and $\Delta_M\neq 0$ phase. But in
all three cases, we recover the normal phase at a high chemical
potential, $S_1 : \mu \ge 8.5,~ S_2 :\mu \ge 12.3 ,~S_3 :\mu \ge
9.7 $. The phase diagram with the parameter set $S_1$ is shown in
fig. \ref{coe}(a). To discuss a possible realization of ISB at the
critical potential $\mu_c$, we note that in the mixed phase $U(1)$
symmetry of the IT model is spontaneously broken to $Z_2$ by the
superconducting gap $\Delta_B$. Thus one may expect an ISB
phenomenon to take place at the critical chemical potential but
one should note that chemical potential does not break $U(1)$
symmetry explicitly at the level of an action. One of the
essential points of ISB is that a small explicit breaking of a
symmetry by chemical potential or external magnetic field at the
level of Lagranigan leads to drastic changes in the vacuum
structure of the theory and thereby gives a jump in number density
\cite{lr} or magnetization \cite{langfeld}. In addition, the mixed
phase in the IT model does not possess the exclusive feature that
the presence of the ISB expel color superconductivity found in
ref. \cite{lr}. Thus we are led to suggest that
 the ISB phenomena may not be a relevant notion in the IT model
\footnote{We are grateful to K. Langfeld for emphasizing this point to us.}.

\subsection{ $\Delta_B\neq 0$, $\lambda_B > 0 $}
\begin{figure}[htb]
\centerline{\epsfig{file=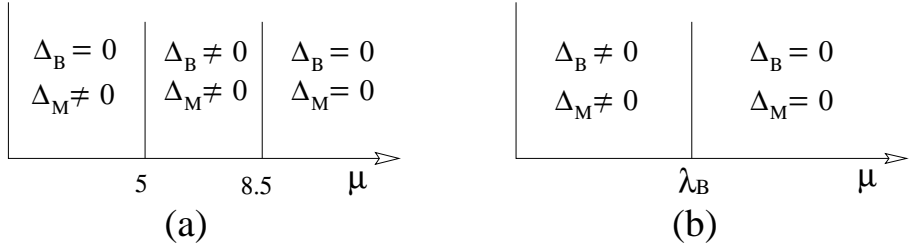,width=12cm}}
\caption{\small The phase diagram calculated in IT model in cold dense matter. Here
energy dimension of chemical potential is arbitrary. }\label{coe}
\end{figure}

 \indent\indent
In this case, the solutions for $\Delta_B$ and $\Delta_M$ are the
same as those in (\ref{sols}) but $E=\lambda_B$
with a restriction $\lambda_B+\mu+2\lambda_M >0$ in the limit
${\beta \rightarrow \infty}$. We find from the ground state energy
given by the parameter choice (\ref{cc1}) that as long as $\mu
<\mu_c =\lambda_B$, we have a mixed phase, $\Delta_B\neq 0$ and
$\Delta_M\neq 0$, and that in the regime $\mu >\mu_c $, our system
sits in the normal phase.
The resulting phase diagram is depicted in fig. \ref{coe} (b).

\section{Field Theory Model For Competitions Between $\la\bar\psi\psi\ra$,
$\la\psi\psi\ra$ And $\la\psi^\dagger\psi\ra$}
 \indent\indent
  In this section, we shall study a field-theory model by generalizing
the model considered recently by Chodos et al \cite{chodos} to one
that contains a number density field corresponding to
$\la\psi^\dagger\psi\ra$. The (1+1)-dimensional toy model we shall
consider is defined by the Lagrangian
\ba
L&=&i\bar\psi^{(i)}\not\!\partial \psi^{(i)}-\mu\psi^\dagger\psi
+\frac{1}{2}g^2[\bar\psi^{(i)} \psi^{(i)}][\bar\psi^{(j)}
\psi^{(j)}]\no
 &&+2G^2[\bar\psi^{(i)}\gamma_5 \psi^{(j)}][\bar\psi^{(i)}\gamma_5 \psi^{(j)}]
-2F^2[\bar\psi^{(i)}\gamma^\mu
\psi^{(i)}][\bar\psi^{(j)}\gamma_\mu \psi^{(j)}]
\ea
where $i$
runs from 1 to $N$. The last term is the term we have added to the
model of Chodos et al. We shall do a large $N$ approximation. We
do a Hubbard-Stratanovich transformation using the auxiliary
fields, $\phi$, $B$ and $V_\mu$ by adding the term
\ba
\Delta
L&=&-\frac{1}{2g^2}[\phi+g^2\bar\psi^{(i)}\psi^{(i)}]^2-\frac{1}{G^2}
(B^\dagger-G^2\epsilon_{\alpha\beta}\psi_\alpha^{\dagger(i)}
\psi_\beta^{\dagger(i)})
(B+G^2\epsilon_{\gamma\delta}\psi_\gamma^{(i)}
\psi_\delta^{(i)})\no &&+\frac{2}{F^2}[V_\mu
+F^2\bar\psi^{(i)}\gamma_\mu\psi^{(i)}]^2.\label{dl}
\ea
 The resulting Lagrangian $L^\prime=L+\Delta L$ is
\ba
L^\prime &=&
\bar\psi^{(i)} (i\not\!\partial -\phi-\mu\gamma_0 +4\not\!
V)\psi^{(i)} -\frac{\phi^2}{2g^2}-\frac{B^\dagger B}{G^2}
+\frac{2}{F^2}V_\mu V^\mu\no & &+B\epsilon_{\alpha\beta}
\psi_\alpha^{\dagger (i)}\psi_\beta^{\dagger (i)} -B^\dagger
\epsilon_{\alpha\beta} \psi_\alpha^{ (i)}\psi_\beta^{ (i)}.
\ea
 We
can now integrate out $\psi$ and $\psi^\dagger$ to obtain the
effective action (modulo a constant)
\ba
\Gamma_{eff}(\phi, B,
B^\dagger, V_0)&=&\int d^2 x (-\frac{\phi^2}{2g^2}-\frac{B^\dagger
B}{G^2} +\frac{2}{F^2}V_0^2 ) -\frac{i}{2}\Tr \ln( A^T A)\no
&&-\frac{i}{2}\Tr \ln[1+M^2(A^T)^{-1}\sigma_2A^{-1}\sigma_2 ]
\ea
where we have assumed that $B$, $B^\dagger$, $\phi$ and $V_0$ are
constant since we are interested in an
effective potential with translation invariance and have defined
\ba
M^2 &=& 4B^\dagger B \no A&=&i\partial_0
+i\sigma_3\partial_3-\tilde\mu-\phi\sigma_1\no
A^T &=&-i\partial_0
-i\sigma_3\partial_3-\tilde\mu-\phi\sigma_1\nonumber
\ea
with $\tilde\mu=\mu-4V_0$.

Factoring out $N$ coming from the flavor trace, we write $g^2
N=\lambda$, $G^2N=\frac{\kappa}{4}$ and $F^2 N=2\alpha$ and define
$V_{eff}$ as $\Gamma_{eff}=-N(\int d^2x) V_{eff}$:
\ba
V_{eff}=
\frac{\phi^2}{2\lambda}+\frac{M^2}{\kappa} -\frac{V_0^2}{\alpha}
+ V_{eff}^{(1)}
\ea
where  $V_{eff}^{(1)} =
\frac{i}{2}[\Tr \ln( A^T A)
+\Tr \ln(1+M^2(A^T)^{-1}\sigma_2A^{-1}\sigma_2 )]$.
 Then the gap equations are obtained by
\ba \frac{\partial V_{eff}}{\partial \phi^2}= \frac{\partial
V_{eff}}{\partial M^2}= \frac{\partial V_{eff}}{\partial V_0^2}=0
\ea from which we get \ba \frac{1}{2\lambda}&=&-\frac{\partial
V_{eff}^{(1)}}{\partial \phi^2}= i\int\frac{d^2k}{(2\pi)^2}
\frac{[k_0^2-k_1^2+\tilde\mu^2 +M^2-\phi^2]}{D} \no
\frac{1}{\kappa}&=&-\frac{\partial V_{eff}^{(1)}}{\partial M^2}=
i\int\frac{d^2k}{(2\pi)^2} \frac{[k_0^2-k_1^2-\tilde\mu^2
-M^2+\phi^2]}{D} \no \frac{1}{\alpha}&=& \frac{\partial
V_{eff}^{(1)}}{\partial V_0^2}=i\int\frac{d^2k}{(2\pi)^2}
\frac{4\tilde\mu[k_0^2+k_1^2-\tilde\mu^2 -M^2+\phi^2]} {V_0 D} \ea
where $D= [k_0^2-k_1^2-\tilde\mu^2 -M^2+\phi^2]^2-4[\phi^2k_0^2 +
\mu^2 k_1^2-\phi^2k_1^2]$ with $k_0=k_0 +i\epsilon\ {\rm sgn}
k_0$. After the $k_0$ integral, we have \ba
\frac{1}{2\lambda}&=&\frac{1}{8\pi}\int_{-\Lambda}^{\Lambda} dk_1
[\frac{1}{k_+} +\frac{1}{k_-}
+\frac{M^2+\tilde\mu^2}{\sqrt{M^2\phi^2 +
\tilde\mu^2(k_1^2+\phi^2)}}(\frac{1}{k_+} -\frac{1}{k_-})]\no
\frac{1}{\kappa}&=&\frac{1}{8\pi}\int_{-\Lambda}^{\Lambda} dk_1
[\frac{1}{k_+} +\frac{1}{k_-} +\frac{\phi^2}{\sqrt{M^2\phi^2 +
\tilde\mu^2(k_1^2+\phi^2)}}(\frac{1}{k_+} -\frac{1}{k_-})]\no
\frac{1}{\alpha}&=&\frac{\tilde\mu}{V_0}
\frac{1}{2\pi}\int_{-\Lambda}^{\Lambda} dk_1 [\frac{1}{k_+}
+\frac{1}{k_-} +\frac{\phi^2+k_1^2}{\sqrt{M^2\phi^2 +
\tilde\mu^2(k_1^2+\phi^2)}}(\frac{1}{k_+}
-\frac{1}{k_-})]\nonumber \ea where $\Lambda$ is a cutoff to
regularize logarithmic divergences\\ and
$k_{\pm}=\sqrt{M^2+\phi^2+\tilde\mu^2+k_1^2\pm 2
\sqrt{M^2\phi^2+\tilde\mu^2(k_1^2+\phi^2)}}$.

By integrating with respect to $\phi^2$, $M^2$ and $V_0^2$, we
have the unrenormalized $V_{eff}^{(1)}$,
\ba
V_{eff}^{(1)}=-\frac{1}{2\pi}\int_{0}^{\Lambda} dk_1 [k_+ - k_-
+c(k_1) ] +C(\mu)
\ea
where $c(k_1)$ and $C(\mu)$, the value of which are
determined later, are constants of integration.
 In the unrenormalized effective potential
in free space, there can be quadratic and logarithmic divergences.
Since free-space renormalization will eliminate all deivergences
even in medium, we will cancel out the quadractic diverence by
choosing a suitable value of $c(k_1)$ and define renormalized
coupling constants to eliminate logarithmic divergences. To be
explicit, we take $\mu=0$. Then the $V_{eff}^{(1)}$ is given by
\ba V_{eff}^{(1)}&=&-\frac{1}{2\pi}\int_0^\Lambda dk_1[
\sqrt{M^2+\phi^2+16V_0^2+k_1^2+ 2
\sqrt{M^2\phi^2+16V_0^2(k_1^2+\phi^2)}}\no
&&+\sqrt{M^2+\phi^2+16V_0^2+k_1^2- 2
\sqrt{M^2\phi^2+16V_0^2(k_1^2+\phi^2)}}-2k_1]\label{eff1} \ea
where we have {\it chosen} $c(k_1)=-2k_1$ to cancel the quadratic
divergence, see (\ref{kke1}) and $C(\mu=0)=0$, see (\ref{kke2}).
\subsection{$\phi=0,~\mu =0$}
 \indent\indent
Since it is not so easy to integrate  (\ref{eff1}) explicitly to
separate the infinities from finite quantities, let us first
confine ourselves to the case of $\phi=0$ \footnote{In Ref.
\cite{chodos}, it is shown that although a solution to the gap
equations exists in which both condensates, $\la\bar\psi\psi\ra$
and $\la\psi\psi\ra$,  are non-vanishing, the global minimum of
the effective potential always occurs for the case when one or the
other condensate vanishes in free space ($\mu=0$) except for one
very special case which we will not consider here. If we take
naively this result, setting $\phi=0$ is not so bad an
approximation. }. The $V_{eff}^{(1)}$ is given by \ba
V_{eff}^{(1)} (V_0,M)&=&-\frac{1}{2\pi}\int_0^\Lambda dk_1[
\sqrt{M^2+ (4V_0 +k_1)^2}+ \sqrt{M^2+ (4V_0-k_1)^2}-2k_1]\no
&=&-\frac{1}{4\pi}[(\Lambda +4V_0)\sqrt{(\Lambda +4V_0)^2+M^2}
+(\Lambda -4V_0)\sqrt{(\Lambda -4V_0)^2+M^2}\no &&+M^2\ln
\frac{\Lambda +4V_0+\sqrt{(\Lambda +4V_0)^2+M^2}}
{4V_0+\sqrt{16V_0^2+M^2}}\no &&+M^2\ln \frac{\Lambda
-4V_0+\sqrt{(\Lambda -4V_0)^2+M^2}}
{-4V_0+\sqrt{16V_0^2+M^2}}-2\Lambda^2]. \label{kke1} \ea The
unrenormalized effective potential  is given by \ba
V_{eff}(V_0,M)&=&M^2(\frac{1}{\kappa} -\frac{1}{4\pi}) -V_0^2
(\frac{1}{\alpha} +\frac{8}{\pi})\no &&-\frac{1}{4\pi}[ M^2\ln
\frac{\Lambda +4V_0+\sqrt{(\Lambda +4V_0)^2+M^2}}
{4V_0+\sqrt{16V_0^2+M^2}}\no &&+M^2\ln \frac{\Lambda
-4V_0+\sqrt{(\Lambda -4V_0)^2+M^2}} {-4V_0+\sqrt{16V_0^2+M^2}}].
\ea We define the renormalized couplings $\kappa_R$ and $\alpha_R$
as \ba \frac{\partial^2 V_{eff}}{\partial B\partial
B^\dagger}|_{M=M_0 ,V_0 =v_0} =\frac{4}{\kappa_R} ,~~~
\frac{\partial^2 V_{eff}} {\partial V_0^2}|_{M=M_0 ,V_0 =v_0}
=-\frac{2}{\alpha_R} \ea and get \ba \frac{1}{\alpha_R}&=&
\frac{1}{\alpha}+\frac{8}{\pi}\no
\frac{1}{\kappa_R}&=&\frac{1}{\kappa}-\frac{1}{4\pi} [\ln
\frac{\Lambda +4v_0+\sqrt{(\Lambda +4v_0)^2+M_0^2}}
{4v_0+\sqrt{16v_0^2+M_0^2}}\no && +\ln \frac{\Lambda
-4v_0+\sqrt{(\Lambda -4v_0)^2+M_0^2}} {-4v_0+\sqrt{16v_0^2+M_0^2}}
-1 ]. \ea The renormalized effective potential is now found to be
\ba V_{eff}^R=M^2(\frac{1}{\kappa_R} -\frac{1}{2\pi}) -V_0^2
\frac{1}{\alpha_R} +\frac{M^2}{4\pi}\ln \frac{M^2}{M_0^2}. \ea
Note that $V_0$ and $M$ are not coupled to each other. Solving the
gap equations for $V_0$ and $M$, we obtain \ba V_0&=&0\no
M^2&=&M_0^2 e^{1-\frac{4\pi}{\kappa_R}}~~{\rm or}~~0. \ea At the
solutions, the effective potential becomes \ba V_{eff}^{R} &=&0
~{\rm for} ~V_0=0 ,~M^2=0\no V_{eff}^{R}
&=&-\frac{M_0^2}{4\pi}e^{1-\frac{4\pi}{\kappa_R}} ~{\rm for}
~V_0=0 ,~M^2\neq 0 .\ea So the phase with  $V_0=0 ,~M^2\neq 0$ is
energetically favored and is a global minimum when $\alpha_R <0$.

\subsection{$\phi=0$,~$\mu\neq 0$}
 \indent\indent
In this case, the renormalized effective potential is given (up to
a constant) by
\ba
 V_{eff}^{R}=M^2(\frac{1}{\kappa_R}
-\frac{1}{2\pi}) -V_0^2 \frac{1}{\alpha_R} -\frac{1}{2\pi}\mu^2
+\frac{4}{\pi}\mu V_0 +\frac{M^2}{4\pi}\ln\frac{M^2}{M_0^2} +
C(\mu) \label{kke2}
\ea
where $C(\mu)$ will be fixed by the condition that
$V_{eff}^R(M=0,V_0=0)=0$. We will make use of the condition
$V_{eff}^R(M=0,V_0=0)=0$ throughout this paper. Then we have the
renormalized effective potential, \ba
V_{eff}^{R}=M^2(\frac{1}{\kappa_R} -\frac{1}{2\pi}) -V_0^2
\frac{1}{\alpha_R} +\frac{4}{\pi}\mu V_0
+\frac{M^2}{4\pi}\ln\frac{M^2}{M_0^2} \label{v1}.
\ea
As far as
$M^2$ is concerned, our effective potential is the same as that
in Ref. \cite{chodos}. Note that since $V_0$ and $M$ do not couple,
it is easy to find solutions of the gap equations. We find \ba
V_0&=&2\frac{\alpha_R}{\pi}\mu ,~~M^2 = 0 ~~{\rm or}\\
V_0&=&2\frac{\alpha_R}{\pi}\mu ,~~M^2 =  M_0^2
e^{1-\frac{4\pi}{\kappa_R}}\label{v35}.
\ea
At the solutions, the effective potential becomes
\ba
V_{eff}^{R}
&=&\frac{4}{\pi^2}\alpha_R \mu^2 ~{\rm for}
~V_0=\alpha_R\frac{2}{\pi}\mu ,~M^2=0\no V_{eff}^{R}
&=&\frac{4}{\pi^2}\alpha_R \mu^2
-\frac{M_0^2}{4\pi}e^{1-\frac{4\pi}{\kappa_R}} ~{\rm for}
~V_0=\alpha_R\frac{2}{\pi}\mu ,~M^2 \neq 0.\label{efp1} \ea
From
(\ref{efp1}), we can see that the phase with $M\neq 0$  is
energetically favored regardless of the behavior of $V_0$. Thus we
 expect there could exist a phase where superconductivity and
mean fields coexist in the case of $M_0^2\neq 0$ at finite
density. Since we are expecting superconductivity at high density,
it is plausible that
 we have $M_0^2\neq 0$ with $M_0$ being small at some relevant density.

\subsection{$M=0, \mu= 0$}
 \indent\indent
In this case, $V_{eff}^{(1)}$ takes the form
 \ba
V_{eff}^{(1)}=-\frac{1}{2\pi}\int_0^\Lambda dk_1 [\mid \sqrt{\phi
^2+k_1^2} +4V_0\mid + \mid \sqrt{\phi^2+k_1^2}  - 4V_0 \mid -2k_1
]. \ea We can consider two cases:
 \vskip 0.5cm

 $\bullet$ $|\phi | > 4|V_0|$:\\ In this case,
the effective potential $V_{eff}^{(1)}$ is independent of $V_0$.
Using the exactly same method adopted in the previous section, we
have \ba \frac{1}{\lambda}&=&\frac{1}{\lambda_R} +\frac{1}{\pi}
\ln\frac{\Lambda+\sqrt{\Lambda^2+\phi^2}}{\phi_0} -\frac{1}{\pi}\\
\frac{1}{\alpha} &=& \frac{1}{\alpha_R}\\
V_{eff}^R&=&\phi^2(\frac{1}{2\lambda_R}-\frac{3}{4\pi})-\frac{1}{\alpha_R}V_0^2
 +\frac{\phi^2}{4\pi} \ln\frac{\phi^2}{\phi_0^2} \label{eff2}
\ea
where $\lambda_R$ is defined by
 $\frac{\partial^2 V_{eff}}{\partial\phi^2}\mid_{M=M_0 , \phi=\phi_0}=
\frac{1}{\lambda_R}$. Since $\alpha$ and $\alpha_R$ differ by a constant, we will use, hereafter,
$\alpha$ instead of $\alpha_R$ just for simplicity.
The solutions of the gap equations are given by
\ba
V_0 &=&0,~~\phi=0\\ ~{\rm or}~
V_0 &=&0,~~\phi= \phi_0 e^{1-\frac{\pi}{\lambda_R}}.\label{v11}
\ea
It can be easily seen that the phase with $V_0=0$ and $\phi\neq 0$, at which
 $V_{eff}^R=-\frac{1}{4\pi}\phi_0^2 e^{2-\frac{2\pi}{\lambda_R}}$,
is favored. This comes as no surprise.
 \vskip 0.5cm

$\bullet$ $|\phi | <4|V_0|$:\\ We find \ba
\frac{1}{\alpha}&=&\frac{1}{\alpha_R} -\frac{2}{\pi}\frac{16|v_0|}
{\sqrt{16v_0^2-\phi_0^2}}\no
\frac{1}{\lambda}&=&\frac{1}{\lambda_R}+\frac{1}{\pi}
\ln\frac{\Lambda +\sqrt{\Lambda^2 +\phi_0^2}}{4|v_0|
+\sqrt{16v_0^2 -\phi_0^2}} +\frac{1}{\pi}\frac{\phi_0^2}{16v_0^2
-\phi_0^2 +4|v_0|\sqrt{16v_0^2 -\phi_0^2}}\no
V_{eff}^R&=&\phi^2(\frac{1}{2\lambda_R} -\frac{1}{4\pi})
-\frac{1}{\alpha}V_0^2\no &&+\frac{1}{2\pi}\phi^2 \ln\frac{4|V_0|
+\sqrt{16V_0^2 -\phi^2}}{4|v_0| +\sqrt{16v_0^2 -\phi_0^2}}\no
&&+\frac{1}{2\pi}\phi^2\frac{\phi_0^2} {16v_0^2 -\phi_0^2
+4|v_0|\sqrt{16v_0^2 -\phi_0^2}}
-\frac{2}{\pi}|V_0|\sqrt{16V_0^2-\phi^2}.
 \ea
  Solving the gap equations, we find that the only possible solutions are $V_0 =0$
and $\phi=0$ where $V_{eff}^R=0$. So at zero chemical potential,
the system is again characterized by $V_0 =0$ and $\phi\neq 0$.
\subsection{$M=0, \mu\neq 0$}
 \indent\indent
 $\bullet$ $|\phi |>4|V_0|$:\\
For $\mu < \mid \phi\mid$, the effective potential is equal to
(\ref{eff2}). In this case, our system is characterized by \ba V_0
=0,~~\phi= \phi_0 e^{1-\frac{\pi}{\lambda_R}}.\label{v42} \ea For
$\mu > \mid\phi\mid$, we have \ba
V_{eff}^R&=&\phi^2(\frac{1}{2\lambda_R}-\frac{3}{4\pi}
)-\frac{1}{\alpha} V_0^2
 -\frac{1}{2\pi}\tilde\mu\sqrt{\tilde\mu^2-\phi^2}\no
&& -\frac{1}{2\pi} \phi^2 \ln\frac{\phi_0}{\tilde\mu
+\sqrt{\tilde\mu^2 -\phi^2}} +\frac{1}{2\pi}\mu^2. \ea
 (Recall that $\tilde{\mu}=\mu-4V_0$.)
The possible solutions of the gap equations are calculated to be
\ba
V_0&=& -\frac{1}{\frac{\pi}{2\alpha} -4}
(\mu -\phi_0e^{1-\frac{\pi}{\lambda_R}})\no
\phi^2&=& \tilde\mu^2-\frac{\pi^2}{4\alpha^2}V_0^2.\label{gs1}
\ea
At the solutions of the gap equations, we have
\ba
V_{eff}^R=\frac{1}{4\pi}\mu^2 +(\frac{2}{\pi}-\frac{1}{4\alpha_R})\mu V_0
-\frac{1}{4\pi}(16-\frac{\pi^2}{4\alpha_R^2})V_0^2. \label{vv1}
\ea

$\bullet$ $|\phi |<4|V_0| $:\\ \ba
V_{eff}^R&=&\phi^2(\frac{1}{2\lambda_R}
-\frac{1}{4\pi})-\frac{1}{\alpha}V_0^2\no &&+\frac{1}{2\pi}\phi^2
\ln\frac{\mid\tilde\mu\mid + \sqrt{\tilde\mu^2-\phi^2}} {4v_0
+\sqrt{16v_0^2 -\phi_0^2}} +\frac{1}{2\pi}\mu^2\no
&&+\frac{1}{2\pi}\phi^2\frac{\phi_0^2} {16v_0^2 -\phi_0^2
+4|v_0|\sqrt{16v_0^2 -\phi_0^2}}
-\frac{1}{2\pi}\mid\tilde\mu\mid\sqrt{\tilde\mu^2-\phi^2}. \ea
First, we consider the case of \ba \phi=0~ {\rm and}~
V_0=\frac{1}{\frac{\pi}{2\alpha} +4}\mu \label{gs2}
 \ea
  which is
one of the solutions of the gap equations. In this case, the
effective potential becomes \ba V_{eff}^R
=-\frac{1}{2\pi}(16+\frac{2\pi}{\alpha} )V_0^2 +
\frac{4}{\pi}V_0\mu.\label{vv2}
\ea
For $\tilde\mu >0$, we have
the following solution for the gap equation
\ba
V_0&=&-\frac{1}{\frac{\pi}{2\alpha}-4}[\mu -C_1
e^{-(\frac{\pi}{\lambda_R}+C_2 ) } ]\no \phi^2&=&\tilde
\mu^2-\frac{\pi^2}{4\alpha^2}V_0^2.\label{gs3} \ea In the case of
$\tilde\mu <0$, we have \ba
V_0&=&-\frac{1}{\frac{\pi}{2\alpha}-4}[\mu +C_1
e^{-(\frac{\pi}{\lambda_R}+C_2 ) } ]\no \phi^2&=&\tilde
\mu^2-\frac{\pi^2}{4\alpha^2}V_0^2.\label{gs4} \ea To see which
phase is energetically favorable, we take $\phi_0=m_F$ and
$M_0\approx 0$~\footnote{Since it is not plausible to have
superconductivity in free space, $M_0$ should be zero
\cite{chodos} or very small, if any. We shall assume, however,
that we can have a small but non-zero value of
 $M_0$ at finite density}.

In free space ($\mu=0$), we have two phases:
\ba
V_{eff}^R&=&-\frac{1}{4\pi}m_F^2 e^{2-\frac{2\pi}{\lambda_R}}
~{\rm for}~V_0 = 0~{\rm and}~ \phi\neq 0\no V_{eff}^{R}
&=&-\frac{M_0^2}{4\pi}e^{1-\frac{4\pi}{\kappa_R}} ~{\rm for}~ V_0
= 0~{\rm and}~ M\neq 0.
\ea
Since $M_0$ will be zero or small, our
system sits in the phase with $V_0=0$ and $\phi\neq 0$. In free
space, therefore, we have chiral symmetry breaking but  no
superconductivity.

\begin{figure}[htb]
\centerline{\epsfig{file=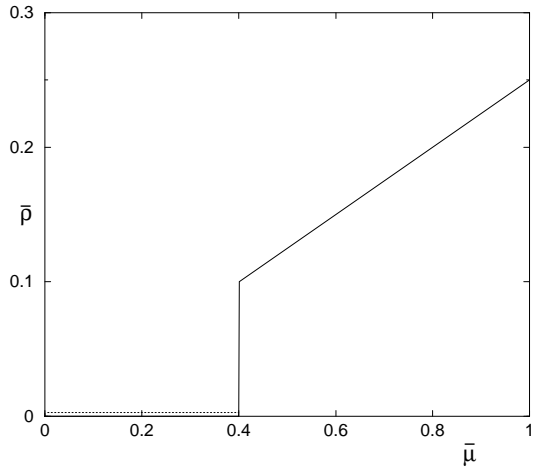,width=7cm}}
\caption{\small  Fermion number density as a function of
chemical potential. Here $\bar\rho \equiv\frac{\rho_F}{m_F}$ and
 $\bar\mu \equiv\frac{\mu}{m_F}$. In the regime
$\mu <\mu_c$ the system sits in the chiral symmetry breaking phase($II$)
and in the mixed phase phase($I$) in the case of $\mu >\mu_c$.  }\label{jump}
\end{figure}

At finite density, the situation becomes more complicated. We have
six sets of solutions of the gap equation: (\ref{v35}),
(\ref{v42}), (\ref{gs1}), (\ref{gs2}), (\ref{gs3}) and
(\ref{gs4}). Now let us investigate the competition between the
phases characterized by (\ref{v35}), (\ref{v42}) and (\ref{gs2}).
The effective potentials given at their ground-state positions are
 \ba I ~:~
V_{eff}^R&=&\frac{4}{\pi^2}\frac{1}{\frac{1}{\alpha}+\frac{8}{\pi}}\mu^2
-\frac{M_0^2}{4\pi^2}e^{1-\frac{4\pi}{\kappa_R}}~\leftrightarrow~
(\ref{v35})\no II~:~V_{eff}^R&=&-\frac{1}{4\pi}m_F^2
e^{2-\frac{2\pi}{\lambda_R}}~\leftrightarrow~ (\ref{v42})\no
III~:~V_{eff}^R&=&\frac{4}{\pi^2}\frac{1}{\frac{1}{\alpha}+\frac{8}{\pi}}\mu^2
~\leftrightarrow~ (\ref{gs2}).
\ea
 Comparing $I$ and $III$, we
find that our system will be
in the mixed phase ($I$) characterized by
$\la\psi^\dagger\psi\ra\neq 0$
and $\la\psi \psi\ra\neq 0$ as long as
$M_0^2\neq 0$. From the condition that the phase
$I$ is the global minimum of the potential (\ref{v1}),
 $\frac{1}{\alpha} +\frac{8}{\pi} <0$,
we find $-0.125\pi <\alpha <0$. It is expected that the phase
$II$, which is valid when $\mu <|\phi |$, becomes the absolute
minimum of the potential at low density. To see the competition
between $I$ and $II$ explicitly, we neglect a term with $M_0^2$ in
the phase I and take $ \lambda_R=\pi$ and $\alpha=-0.1\pi$.
Then we find $\frac{\mu_c}{m_F}\approx 0.4$. In the regime $\mu
<\mu_c$, we have the chiral symmetry breaking phase ($II$) with
almost zero fermion number density \footnote{If we consider
density dependence of $m_F$ and $\lambda_R$, the fermion number
density may not be zero but is still expected to be small.}
defined by $\rho_F =-\frac{\partial V_R^{eff}}
{\partial\tilde\mu}$.
In the case of $\mu >\mu_c$ the system sits
in the mixed phase ($I$)
 with  fermion number density
\ba \rho_F &=&-\frac{8\alpha_R}{\pi^2-8\pi\alpha_R}\mu ~~{\rm
with}~\frac{1}{\alpha_R}= \frac{1}{\alpha} + \frac{8}{\pi}\no
&\approx& 0.25\mu ~{\rm for}~\alpha=-0.1\pi . \ea The fermion
number density is depicted in Fig. \ref{jump}. We should, however,
consider competitions with other possible phases. To investigate
numerically the phases characterized by (\ref{gs1}),  (\ref{gs3})
and (\ref{gs4}), we take the following parameters \ba
\lambda_R&=&\pi\no \alpha &=&-\pi,~-0.1\pi\no
\frac{v_0}{\phi_0}&=&10,~1.0,~0.3~. \ea With this parameter
choice, we plot the renormalized effective potentials near the
minimum-energy positions (i.e., solutions of the gap equations),
(\ref{gs1}),  (\ref{gs3}) and (\ref{gs4}). We find that the phases
characterized by (\ref{gs1}),  (\ref{gs3}) and (\ref{gs4})
correspond to the maximum  or saddle point of the renormalized
effective potential or does not satisfy some constraints, for
example $4|V_0|>|\phi|$, near the minimum-energy positions and
therefore those phases are unstable or an unphysical ``vacuum."

\section{Conclusion}
 \indent\indent
We studied the competitions between induced symmetry breaking (ISB),
Cooper pairing condensate
 and chiral condensate at finite baryon density.

By reformulating the IT model~\cite{it1} using the
Bogoliubov-Valatin transformation, we show that the mixed phase,
in which both $\Delta_M$ and $\Delta_B$ are non-zero, is
energetically favored. By symmetry we argue that the ISB phenomena
may not be a relevant notion in the IT model.

We calculated the effective potentials $V(\phi=0, V_0, M)$ and
$V(M=0, V_0, \phi)$ which are functions of the fermion-antifermion
(or chiral), ISB (or mean-field) and Cooper-pairing order
parameters $\phi$, $V_0$ and $M$, respectively. At zero chemical
potential, we find that the global minimum of the effective
potential is given by $V_0=0$ and $M\neq 0$ or $V_0=0$ and
$\phi\neq 0$. Since $M_0$ should be zero \cite{chodos} or very
small, it is reasonable to conclude
that the system sits in the chiral symmetry breaking phase.

At finite density, taking $\lambda_R=\pi$ and $\alpha =-0.1\pi$,
we find that in the regime $\mu <\mu_c$, we have the chiral symmetry breaking
phase(II) with almost zero fermion number density and in the case
of $\mu >\mu_c$ the system sits in the mixed phase with nonzero
``vacuum" (or rather ground-state) expectation values of
$\la\psi^\dagger\psi\ra$ and $\la\psi\psi\ra$, {\it i.e}, $V_0\neq
0$ and $M\neq 0$. We observe an ISB-like behavior in the fermion
number density as in fig. \ref{jump} but we fail to
observe the exclusive competition obtained in \cite{lr}.
As we can see in (\ref{v35}) and also in
ref.\cite{chodos}, $M$ ($\sim\la\psi\psi\ra $) is independent of the
chemical potential. We are unable to say whether or not these are
an artifact of the simplified models not present in effective theories
of QCD. A similar (uncertain) situation applies also to analysis made in QCD
at weak coupling\cite{pisa}.

Our analysis depends on the value of the renomalization point
which is arbitrary in general.  The possible resolution to this
arbitrariness is to do a renormalization group analysis at
finite density discussed in \cite{kim}. Such a procedure will
replace the renormalization scale (up to a constant) by the chemical
potential\cite{kim}. We leave this exercise to a future
publication.

\section*{Acknowledgment}
\indent\indent
Part of this work was done at Korea Institute of Advanced Studies,
Seoul, Korea and at the Theory Group, State University of New York,
Stony Brook, N.Y. The hospitality of the two institutes is
highly appreciated. We thank Kurt Langfeld for helpful comments.
YK is supported in part by the Korea Ministry of Education (BSRI
99-2441) and KOSEF (Grant No. 985-0200-001-2) and he thanks Prof. Hyun Kyu Lee
for his support.

\section*{Appendix}
\indent\indent In this appendix, we compare the grand canonical
ensemble used in this work and in ref.\cite{it1,it2} (case {\bf
B}) with that widely used in BCS theory, for example see
\cite{blatt} (case {\bf A}). Here ($a,a^*$) stand for the
annihilation and creation operators for the ``bare" fermions and
($b,b^*$) for the quasiparticles as in section 2.

\vskip 0.7cm
{\underline{$\Delta_B\neq 0 ,~\Delta_M=0$}}

\indent
 Here we shall
take the mean-field gap to be zero, i.e., $\Delta_M =0$ for
simplicity. Since the BCS ground state does not preserve the
particle number of the ground state, a condition is imposed  on
the average; \ba \la\Psi_{BCS} | N_{op} |\Psi_{BCS} \ra =\bar N
\ea where $ N_{op} =\sum_{k} a^*_k a_k$, the number operator. As
described in ref.\cite{blatt} in detail, this relaxation of the
condition of $N$ in the  BCS and Bogoliubov theory  must be
distinguished from the use of grand canonical ensemble in
statistical mechanics.
 In grand canonical ensemble of statistical mechanics,
we deal with an ensemble of systems, with a distribution of
particle numbers $N$ with the systems with N particles being
weighted by a factor $z^N$ where $z$ called fugacity is defined by
$z=e^{\beta\mu}$. However, each separate system within an assembly
has its own definite number of particles $N$.

In the case of {\bf A}, we add a term $-\mu_a a^* a$ to the
Hamiltonian to specify the mean value of particle number.
 Normally in this case, we  adjust the chemical potential $\mu$ to satisfy
$\la N_a\ra ={\bar N}_a$. Here the chemical potential $\mu_{a}$ is
a variable conjugate to $ N_{a} =\sum_{k} a^*_k a_k $. After
specifying the mean number of $ N_{a}$, we diagonalize the
Hamiltonian to get $H (b^* b)$ describing quasi-particles. We can
specify a constant number (in the sense of the average given
above) to a system characterized by this diagonalized Hamiltonian
$H (b^* b)$. Therefore this system becomes {\it effectively}
canonical ensemble of the statistical system with definite number
but  we are {\it essentially} using the grand canonical ensemble.
Then the probability of quasi-particle thermal excitation is given
at a given energy $\ep$ by
 $\la b^*b\ra =(1+ e^{\beta \sqrt{(\ep -\mu_{a} )^2 +\Delta_B^2} })^{-1}$.

In the case of {\bf B},
 we do not constrain the mean number in terms of $ N_{a} =\sum_{k} a^*_k a_k $.
Since our diagonalized Hamiltonian $H (b^* b)$ has the same
structure with that of non-interacting harmonic oscillator and
preserves the particle number, we can define a {\it simultaneous}
eigenstate of quasi-particle number and energy
 operators
\footnote{This is not true in the case of BCS and Bogoliubov
theory.}. Note that the quasiparticle operator $b$ kills, by
definition, the BCS ground state $ b |\Psi_{BCS}\ra =0 $ and the
lowest quasiparticle excitation state is given by $ b^* |\Psi_{BCS}\ra
$. Following an elementary procedure in statistical mechanics, we
have $\la b^*b\ra =(e^{\beta ( \sqrt{\epsilon^2 +\Delta_B^2}
-\mu_b)} +1)^{-1}$ with $\mu_b$ being the ``chemical potential"
for quasiparticles. This formalism {\bf B} is essential if we want
to consider both $\lambda_B >0$ and  $\lambda_B <0$ on the same
footing. If we use the formalism {\bf A}, the gap equation becomes
\ba \sqrt{(\epsilon -\mu_{a})^2 +\Delta_B^2} =\lambda_B \tanh
\frac{\beta \sqrt{(\epsilon -\mu_{a})^2 +\Delta_B^2}}{2}. \ea
Taking $\beta\rightarrow\infty$, we have \ba \sqrt{(\epsilon
-\mu_{a})^2 +\Delta_B^2} =\lambda_B \ea showing that $\lambda_B
<0$ does not enter into the consideration. Note that  $\lambda_B
>0$ corresponds to the usual BCS attraction.

Since the chemical potential $\mu_b$ essentially controls the
number of quasiparticle excitations while  the chemical potential
$\mu_{a}$ fixes the mean number of the ground state,
 one might say that $\mu_{b}$ has nothing to do with $\mu_a$.
{\it But} one should keep it in mind that $b^*$ is a linear combination of
 the $a$ and $a^*$ original fermion operators. Thus
$b^*$ can create and annihilate fermions ($a,a^*$) from the ground
state. This implies that controlling the number of quasi-particles
must have something to do with that of the ``bare" fermion $(a^*,
a)$. Whether we fix the number of ground state($A$) or
quasi-particle excitation($B$), they should give the same answer
for $\la a^* a\ra$. Note that $(a^*, a)$ and $(b^*,b)$ are
connected by the canonical transformation:
 \ba
  \la a^*(p)a(p)\ra &=&
\{\frac{c^2(p)}{1+e^{\beta (E-\mu_b)}} +
\frac{s^2(p)}{1+e^{-\beta (E-\mu_b)} } \} ~{\rm
with}~E=\sqrt{\ep^2 +\Delta_B^2}\no &=&
\{\frac{c^2(p)}{1+e^{\beta E}} + \frac{s^2(p)}{1+e^{-\beta
E} } \} ~{\rm with}~E =\sqrt{(\epsilon -\mu_{a})^2 +\Delta_B^2}
 \ea
  where  $\epsilon =p^2$.

\vskip 0.7cm
{\underline{$\Delta_B =0 ,~\Delta_M\neq 0$}}

\indent
In this case the probability of thermal excitation of ``bare'' fermion is
given by (for $\ep + \Delta_M >\mu_a$)
\ba
  \la a^*(p)a(p)\ra &=&
\frac{1}{1+e^{\beta (\ep + \Delta_M -\mu_b)}}  ~{\rm
with}~B\no
&=&\frac{1}{1+e^{\beta (\ep + \Delta_M -\mu_a)}}
 ~{\rm with}~A\no
&=& \la b^*(p)b(p)\ra
\ea
showing $\mu_a =\mu_b$. Note that in the case
$\Delta_B =0 ,~\Delta_M\neq 0$, the KMS-state is defined for the operator
$(a,a^*)$\cite{it1}.

\end{document}